# Conceptual difficulties in the use of statistical inference in citation analysis

Ludo Waltman

Centre for Science and Technology Studies, Leiden University, The Netherlands
waltmanlr@cwts.leidenuniv.nl

In this comment, I discuss the use of statistical inference in citation analysis. In a recent paper, Williams and Bornmann argue in favor of the use of statistical inference in citation analysis. I present a critical analysis of their arguments and of similar arguments provided elsewhere in the literature. My conclusion is that the use of statistical inference in citation analysis involves major conceptual difficulties and, consequently, that the usefulness of statistical inference in citation analysis is highly questionable.

## 1. Introduction

Citation-based indicators are sometimes considered to be subject to randomness or chance. This is for instance motivated by the idea that some citations seem to have been carefully chosen by the authors of the citing publication while other citations seem to have been chosen in a more arbitrary or coincidental way. Williams and Bornmann (2016) provide the following illustration of this idea: "How often a paper or collection of papers gets cited might be affected by how many people chose to read a particular issue of a journal or who happened to learn about a paper because somebody casually mentioned it to them". Arbitrariness in the choice of citations was already discussed by Dieks and Chang (1976), who argue that "authors who are giving references always have to choose from a number of considered papers; whether a given paper is cited or not depends on all kinds of personal factors, and this introduces a random element in the total number of citations". The general idea of citation-based indicators being subject to randomness or chance seems to be widely accepted, and I have also adopted this perspective myself (Hicks et al., 2015; Waltman, Van Eck, & Wouters, 2013).

If one considers citation-based indicators to be subject to randomness, a next step could be to move beyond a purely descriptive approach to citation analysis and to introduce the use of statistical inference, for instance significance tests and confidence intervals. In a statistical inference approach to citation analysis, the idea is to formally model the effect of randomness on citation-based indicators and to quantify the resulting uncertainty in these indicators. Statistical inference has been used quite commonly in the literature on citation analysis. Examples include early work by Dieks and Chang (1976) and Schubert and Glänzel (1983) and more recent work by Opthof and Leydesdorff (2010), Stern (2013), Abramo, D'Angelo, and Grilli (2015, 2016), Fairclough and Thelwall (2015), Thelwall (2016), and Williams and Bornmann (2014, 2016). This comment relates mainly to the work by Williams and Bornmann (2016; henceforth WB), although I will also briefly consider some other papers in which statistical inference is used in citation analysis.

My aim in this comment is to make clear that the use of statistical inference in citation analysis involves major conceptual difficulties. To some degree, WB indeed recognize these conceptual difficulties. WB point out that in citation analysis one often has available data on essentially all publications, and the corresponding citations, of a research institution (at least all publications and citations within the universe of a specific bibliographic database, such as Web of Science or Scopus). One then seems to have access to the entire population rather than just a sample drawn from a population. In the terminology of Berk, Western, and Weiss (1995), the available data represents an apparent population. So far, this issue has hardly been discussed in the literature on citation analysis (for an exception, see Schneider, 2013). It is laudable that WB present an explicit argument that aims to justify the use of statistical inference in citation analysis even when the available data represents an apparent population.

However, as I will make clear in this comment, the conceptual difficulties go much further than suggested by WB. I will argue that there is no objective notion of randomness in citation analysis. What is seen as random and what is not depends on the perspective that is taken and therefore is subjective. In addition, I will also argue that, when statistical inference is used in citation analysis, it is crucial to be explicit about the type of randomness that is considered. WB are not very clear about this, and the type of randomness based on which they motivate their use of statistical inference appears to be different from the type of randomness on which their statistical modeling approach is actually focused.

## 2. Subjectivity of randomness in citation analysis

In citation analysis, which factors are seen as random and which are not depends on how exactly the idea of randomness is conceived. There is no objective concept of randomness in citation analysis.[1] Instead, one needs to subjectively decide which factors are seen as random and which are not. For instance, WB seem to regard a citation to a publication as random if the citing authors "happened to learn about a paper because somebody casually mentioned it to them". However, when authors cite a publication, they always must have learned about the publication in some way. The authors for instance may have found the publication by browsing through a journal issue, by performing a literature search using Google Scholar, by attending a conference presentation, by following a mailing list, by reading a tweet, or indeed by casually being informed about the publication by someone else. Which of these ways of learning about a publication should be seen as random, and which are non-random? There is no clear answer to this question. What is seen as random and what is not depends on the perspective that is taken and therefore is subjective.

To motivate the use of statistical inference when one has available data on all publications of a research institution, WB suggest the hypothetical possibility of citation processes being repeated multiple times: "If we could somehow repeat the citation process over and over ..., the citation impact of papers ... would not be exactly the same for each repetition". This idea of citation processes being repeated multiple times does not provide an objective concept of randomness either. The difficulty is that one needs to decide which factors are allowed to change in different repetitions of a citation process and which factors are treated as fixed. Indeed, as suggested by WB, in the hypothetical situation in which a citation process is repeated, one could allow for the possibility that a publication ends up in a different issue of a journal and, consequently, attracts more or less attention and receives a larger or smaller number of citations. Of course, one could easily think of many other possibilities as well. When a citation process is repeated, perhaps a publication will address a completely different topic and will therefore receive a larger or smaller number of citations. Do we allow the topic of a publication to change when repeating a citation process? Do we allow the peer review process of a publication to have a different outcome? Do we allow a publication to appear in a different journal? Do we allow for the possibility that an entire research community loses interest in a particular topic and therefore does not cite anymore publications dealing with this topic? If one accepts the hypothetical idea of citation processes being repeated multiple times, one still needs to make a subjective decision on the factors that are allowed to change and the factors that are treated as fixed.

---

[1] An exception could be the situation, also mentioned by WB, in which one has available data only for a sample of the publications of a research institution, not for all publications, and in which one is interested in making statements about all publications (at least all publications within the universe of a specific bibliographic database). However, this situation is uncommon. In citation analysis, one usually has available data on all publications of a research institution.

WB do not recognize the subjective nature of the concept of randomness in citation analysis. They seem to suggest the existence of a clear dichotomy between random and non-random factors, but they do not explain how exactly the distinction between random and non-random factors can be made. Although WB provide a few examples of what they consider to be random factors, they do not characterize the concept of randomness in citation analysis in a systematic way. It should be noted that other papers in which statistical inference is used in citation analysis suffer from the same problem. Many papers do not provide any discussion at all of the concept of randomness in citation analysis. A few papers (Abramo et al., 2015, 2016; Dieks & Chang, 1976) do provide some discussion, but they fail to provide a clear and unambiguous description of the concept of randomness. This supports the idea that there is no objective concept of randomness in citation analysis. What is seen as random and what is not is a subjective decision.

## 3. Different perspectives on randomness in citation analysis

To simplify things a bit, I would like to distinguish between a few main perspectives on randomness in citation analysis. Suppose one applies citation-based indicators at the level of a research institution, and suppose one has the idea that the indicators are influenced by randomness. Before the use of statistical inference can be considered, one then first needs to be more explicit on the concept of randomness that one has in mind. As pointed out above, there is no objective concept of randomness in citation analysis. Instead, one can take many different perspectives on randomness in citation analysis. There seem to be three main perspectives, which for simplicity I refer to as type 1, type 2, and type 3 randomness. The general idea of these three perspectives can be summarized as follows:

- *Type 1 randomness: Randomness at the level of the citations received by a publication*

    When researchers choose the references they include in their publications, they sometimes overlook relevant work and they sometimes by mistake cite work that is of little or no relevance. Also, when there is a lot of relevant work that in principle could be cited, researchers may in a more or less arbitrary way choose to cite only some of this work. Because of arbitrariness in the choice of references, the number of citations received by a publication may be considered subject to randomness.

- *Type 2 randomness: Randomness at the level of the publications produced by a researcher (or a research institution)*

    A researcher (or a research institution) typically produces both publications of higher quality and publications of lower quality. On average, higher-quality publications can be expected to receive more citations than lower-quality publications. Incidental fluctuations in the quality of the publications produced by a researcher (or a research institution) cause fluctuations in citation-based statistics. One might want to regard these fluctuations as randomness.

- *Type 3 randomness: Randomness at the level of the researchers employed by a research institution*

    A research institution typically employs both more talented researchers and less talented researchers. On average, publications produced by more talented researchers can be expected to receive more citations than publications produced by less talented researchers. Incidental fluctuations in the talent of the researchers employed by a research institution cause fluctuations in citation-based statistics. One might want to regard these fluctuations as randomness.

Which type of randomness is considered by WB? This is not very clear. WB seem to use type 1 randomness to motivate their use of statistical inference. They state that "how often a paper or collection of papers gets cited might be affected by how many people chose to read a particular issue of a journal or who happened to learn about a paper because somebody casually mentioned it to them". Likewise, WB argue that "chance factors could have increased the number of citations a paper received or else decreased them". These statements suggest that WB are interested in type 1 randomness, that is, randomness at the level of the citations received by individual publications.

However, the statistical modeling approach taken by WB gives a different impression. If their statistical modeling approach had indeed focused on type 1 randomness, a statistical model somewhat similar to the one used by Dieks and Chang (1976) would have been needed. In such a model, each publication has a certain expected number of citations, where the expected number of citations can be different for different publications of the same research institution (e.g., because of quality differences between publications), and the actual number of citations of a publication is a random variable with mean equal to the publication's expected number of citations. Yet, in the statistical modeling approach taken by WB, for each publication of a research institution the number of citations is drawn from the same probability distribution. Each publication therefore has the same expected number of citations (or, since WB take a binary perspective and focus on publications being highly cited or not, each publication has the same probability of being highly cited). It is not clear how such a model can be interpreted in terms of type 1 randomness. Instead, the model seems to describe type 2 randomness: On average, the publications of a research institution have a certain quality and, related to this, a certain propensity to receive citations, but because of 'random' differences between publications, for instance in their quality, for individual publications the number of citations is above or below the expected number.

Papers in which statistical inference is used in citation analysis employ different statistical modeling approaches. These different approaches focus on different types of randomness, although unfortunately the authors of the papers typically do not discuss this. The statistical modeling approach taken by Dieks and Chang (1976) considers type 1 randomness, while the approach taken by Opthof and Leydesdorff (2010), Fairclough and Thelwall (2015), and Thelwall (2016) seems to focus on type 2 randomness. Schubert and Glänzel (1983) and Stern (2013) also take an approach that appears to focus on type 2 randomness, but they do so in the context of journals rather than research institutions. The work by Abramo et al. (2015, 2016) is particularly noteworthy. Their statistical modeling approach is unique because of its focus on type 3 randomness. However, the authors do not seem to recognize this themselves. Abramo et al. (2015) provide a quite extensive discussion of randomness in citation analysis, but this discussion is concerned with type 1 and type 2 randomness, not with type 3 randomness.

## 4. Conclusion

The intuitive idea of citation-based indicators being subject to randomness or chance is reasonable. In principle, the use of statistical inference in citation analysis therefore could make sense. However, as I have pointed out in this comment, the use of statistical inference in citation analysis involves major conceptual difficulties. Statistical inference therefore should be used only if two conditions are satisfied. First, one should recognize that there is no objective notion of randomness in citation analysis. What is seen as random and what is not depends on the perspective that one takes and therefore is subjective. Second, one should be very clear about the type of randomness that one is interested in, and one should employ a statistical modeling approach that indeed focuses on this type of randomness. In other words, one should make sure that the informal discussion of randomness in citation analysis and the formal modeling of this randomness are consistent with each other.

These two conditions are highly demanding, and therefore I am skeptical about the usefulness of statistical inference in citation analysis. In practice, it seems likely that the use of statistical inference will lead to confusion and misunderstandings. As I have tried to make clear in this comment, bibliometric researchers have serious problems with the interpretation of analyses that use statistical inference. Giving a correct interpretation to such analyses will be even more challenging for end users of citation-based indicators.

Contrary to WB, my conclusion is that in citation analysis it is best to restrict oneself to the use of descriptive statistics and to avoid using inferential statistics. One could try to quantify uncertainty in citation analysis, but this should not be done within the framework of statistical inference. A possible approach that could be considered is the use of stability intervals (Colliander & Ahlgren, 2011; Waltman et al., 2012). These intervals, which are for instance used in the CWTS Leiden Ranking (www.leidenranking.com), provide some basic insight into the stability or robustness of citation-based indicators.

## Acknowledgements

I would like to thank François Claveau, Tina Nane, Jesper Schneider, and an anonymous referee for their comments on earlier drafts of this paper. I am grateful to the members of the Advanced Bibliometric Methods working group at the Centre for Science and Technology Studies of Leiden University for their feedback on the ideas presented in this paper.